\documentclass{scrartcl}

\makeatletter
\DeclareOldFontCommand{\rm}{\normalfont\rmfamily}{\mathrm}
\DeclareOldFontCommand{\sf}{\normalfont\sffamily}{\mathsf}
\DeclareOldFontCommand{\tt}{\normalfont\ttfamily}{\mathtt}
\DeclareOldFontCommand{\bf}{\normalfont\bfseries}{\mathbf}
\DeclareOldFontCommand{\it}{\normalfont\itshape}{\mathit}
\DeclareOldFontCommand{\sl}{\normalfont\slshape}{\@nomath\sl}
\DeclareOldFontCommand{\sc}{\normalfont\scshape}{\@nomath\sc}
\makeatother

\usepackage{arxiv}
\usepackage{authblk}

\usepackage{amssymb}
\usepackage{lineno}
\usepackage{refcount}
\usepackage[utf8]{inputenc}
\usepackage{booktabs}
\usepackage[export]{adjustbox}
\usepackage{graphicx}
\usepackage{amsmath}
\usepackage{gensymb}
\usepackage{tikz}
\usepackage{tikz-3dplot}
\usepackage{varwidth}
\usepackage{bm}
\usepackage{pgf}
\usepackage{pgfplots}
\usepackage[english]{babel}
\usepackage{pgfplotstable}
\usepackage{mathptmx} 
\pgfplotsset{compat=1.8}
\usepackage{tikz-dimline}
\usetikzlibrary{fit, positioning, shapes, shadows, calc, intersections, through, backgrounds, arrows, matrix, patterns}
\usepackage[utf8]{inputenc}
\usepackage{amsmath}
\usepackage{amsfonts}
\usepackage{amssymb}
\usepackage[obeyDraft, obeyFinal]{todonotes}
\usepackage{bm}
\usepackage{tcolorbox}
\usepackage{pgfplots}
\usepackage{subfiles}
\usepackage{colortbl}
\usepackage{ctable}
\usepackage{etex}
\usepackage{paralist}
\usepackage{wasysym}
\usepackage{url}
\usepackage{hyperref}
\usepackage{subcaption}
\usepackage{multirow}
\usepackage{pifont}
\usepackage{wrapfig}
\usepackage{placeins}
\usepackage{acronym}
\usepackage[%
  backend=bibtex      
 ,style=numeric-comp  
 ,sorting=none        
 ,sortcites=true      
 ,block=none
 ,indexing=false
 ,citereset=none
 ,isbn=true
 ,url=true
 ,urldate=long
 ,doi=true            
 ,natbib=true         
]{biblatex}
\usepackage{csquotes}

\pgfdeclarelayer{bg}    
\pgfdeclarelayer{fg}    
\pgfsetlayers{bg,main,fg}  

\definecolor{participantAcolor}{RGB}{243,98,33}
\definecolor{participantBcolor}{RGB}{0,102,189}
\definecolor{participantCcolor}{RGB}{66,70,50}
\definecolor{cplColor}{RGB}{0,0,0}
\definecolor{preciceColor}{RGB}{ 161, 177, 25}

\usetikzlibrary{shapes.arrows, fadings}
\tikzfading[name=fade right,
  left color=transparent!0, right color=transparent!100]
  \tikzfading[name=fade middle,
  left color=transparent!0, right color=transparent!50]

\definecolor{ci_pantone300}{RGB}{ 0, 101, 189}

\definecolor{ci_pantone301}{RGB}{ 0, 82, 147}
\definecolor{ci_pantone540}{RGB}{ 0, 51, 89}

\definecolor{ci_pantone283}{RGB}{ 152, 198, 234}
\definecolor{ci_pantone542}{RGB}{ 100, 160, 200}
\definecolor{ci_ivory}{RGB}{ 218, 215, 203}
\definecolor{ci_orange}{RGB}{ 227, 114, 34}
\definecolor{ci_green}{RGB}{ 162, 173, 0}

\tikzstyle{doublefullarrow}=[<->, >=triangle 60, thin]
\tikzstyle{mycircle}=[circle, draw=black!70, semithick, text=black]
\tikzset{>=latex}

\makeatletter
\pgfkeys{%
  /tikz/on layer/.code={
    \pgfonlayer{#1}\begingroup
    \aftergroup\endpgfonlayer
    \aftergroup\endgroup
  },
  /tikz/node on layer/.code={
    \gdef\node@@on@layer{%
      \setbox\tikz@tempbox=\hbox\bgroup\pgfonlayer{#1}\unhbox\tikz@tempbox\endpgfonlayer\egroup}
    \aftergroup\node@on@layer
  },
  /tikz/end node on layer/.code={
    \endpgfonlayer\endgroup\endgroup
  }
}
\def\node@on@layer{\aftergroup\node@@on@layer}
\makeatother

\pgfkeys{/pgfplots/tuftelike/.style={
  semithick,
  tick style={major tick length=4pt,semithick,black},
  separate axis lines,
  axis x line*=bottom,
  axis x line shift=10pt,
  axis y line*=left,
  axis y line shift=10pt,
  enlarge x limits=false,
  enlarge y limits=false
  }}
  
\tikzset{add reference/.style={insert path={%
    coordinate [pos=0,xshift=-0.5\pgflinewidth,yshift=-0.5\pgflinewidth] (#1 south west) 
    coordinate [pos=1,xshift=0.5\pgflinewidth,yshift=0.5\pgflinewidth]   (#1 north east)
    coordinate [pos=.5] (#1 center)                        
    (#1 south west |- #1 north east)     coordinate (#1 north west)
    (#1 center     |- #1 north east)     coordinate (#1 north)
    (#1 center     |- #1 south west)     coordinate (#1 south)
    (#1 south west -| #1 north east)     coordinate (#1 south east)
    (#1 center     -| #1 south west)     coordinate (#1 west)
    (#1 center     -| #1 north east)     coordinate (#1 east)   
}}}  

\makeatletter
\pgfdeclareshape{document}{
\inheritsavedanchors[from=rectangle] 
\inheritanchorborder[from=rectangle]
\inheritanchor[from=rectangle]{center}
\inheritanchor[from=rectangle]{north}
\inheritanchor[from=rectangle]{south}
\inheritanchor[from=rectangle]{west}
\inheritanchor[from=rectangle]{east}
\backgroundpath{
\southwest \pgf@xa=\pgf@x \pgf@ya=\pgf@y
\northeast \pgf@xb=\pgf@x \pgf@yb=\pgf@y
\pgf@xc=\pgf@xb \advance\pgf@xc by-10pt 
\pgf@yc=\pgf@yb \advance\pgf@yc by-10pt
\pgfpathmoveto{\pgfpoint{\pgf@xa}{\pgf@ya}}
\pgfpathlineto{\pgfpoint{\pgf@xa}{\pgf@yb}}
\pgfpathlineto{\pgfpoint{\pgf@xc}{\pgf@yb}}
\pgfpathlineto{\pgfpoint{\pgf@xb}{\pgf@yc}}
\pgfpathlineto{\pgfpoint{\pgf@xb}{\pgf@ya}}
\pgfpathclose
\pgfpathmoveto{\pgfpoint{\pgf@xc}{\pgf@yb}}
\pgfpathlineto{\pgfpoint{\pgf@xc}{\pgf@yc}}
\pgfpathlineto{\pgfpoint{\pgf@xb}{\pgf@yc}}
\pgfpathlineto{\pgfpoint{\pgf@xc}{\pgf@yc}}
}
}
\makeatother

\usepackage{tikz}
\usepackage{pgfplots}
\usepackage{amsmath}
\pgfplotsset{compat=1.9}

\pgfkeys{/pgfplots/tuftelike/.style={
  width=10cm,
  height=4cm,
  semithick,
  tick style={major tick length=4pt,semithick,black},
  separate axis lines,
  axis x line*=bottom,
  axis x line shift=10pt,
  xlabel shift=10pt,
  axis y line*=left,
  axis y line shift=10pt,
  ylabel shift=10pt}}
  
\usetikzlibrary{pgfplots.dateplot}
\usepackage{pgfplotstable}
\usepackage{filecontents}

\usepackage{hyperref}

\makeatletter
\define@key{FV}{highlightlines}{\edef\FV@HighlightLinesList{#1}}
\makeatother

\usepackage[frozencache]{minted}
\setminted{linenos, highlightcolor=blue!10, breaklines, fontsize=\small}

\usepackage{tikz}
\usetikzlibrary{calc,tikzmark,decorations.pathreplacing,calligraphy}

\usepackage{paralist}

\usepackage{float}
\usepackage{subcaption}
\restylefloat{table}

\title{FEniCS-preCICE: Coupling FEniCS to other Simulation Software}

\author[1]{Benjamin Rodenberg}
\author[2]{Ishaan Desai}
\author[1]{Richard Hertrich}
\author[3]{Alexander Jaust}
\author[2]{Benjamin Uekermann}

\affil[1]{Scientific Computing in Computer Science, Department of Informatics, Technical University of Munich, Germany, \texttt{benjamin.rodenberg@in.tum.de}}
\affil[2]{Usability and Sustainability of Simulation Software, Institute for Parallel and Distributed Systems, University of Stuttgart, Germany, \texttt{\{ishaan.desai, benjamin.uekermann\}@ipvs.uni-stuttgart.de}}
\affil[3]{Simulation of Large Systems, Institute for Parallel and Distributed Systems, University of Stuttgart, Germany, \texttt{alexander.jaust@ipvs.uni-stuttgart.de}}

\begin{document}

\maketitle

\begin{abstract}
The new software FEniCS-preCICE is a middle software layer, sitting in between the existing finite-element library FEniCS and the coupling library preCICE. 
The middle layer simplifies coupling (existing) FEniCS application codes to other simulation software via preCICE.
To this end, FEniCS-preCICE converts between FEniCS and preCICE mesh and data structures, provides easy-to-use coupling conditions, and manages data checkpointing for implicit coupling.
The new software is a library itself and follows a FEniCS-native style. Only a few lines of additional code are necessary to prepare a FEniCS application code for coupling.
We illustrate the functionality of FEniCS-preCICE by two examples: a FEniCS heat conduction code coupled to OpenFOAM and a FEniCS linear elasticity code coupled to SU2. 
The results of both scenarios are compared with other simulation software showing good agreement.

\end{abstract}


\section{Motivation and significance}
\label{sec:motivation}

Enabling simulations to play a significant role in answering the great research challenges of our time -- let it be nuclear fusion, personalized medicine or climate prediction -- requires the efficient interplay of diverse simulation software \cite{Keyes2013_MultiPhysics}. Ideally, single simulation components may be treated as black boxes and may easily be plugged together or exchanged. The software preCICE \cite{preCICE} helps to do exactly that: It can be used to glue together arbitrary many of such black-box simulation components. In the following, we refer to these components as \emph{participants} of a coupled simulation. preCICE focuses mainly on mesh-based discretizations of PDE models as participants.
The finite element software FEniCS \cite{AlnaesBlechta2015a, LoggMardalEtAl2012} is a popular choice to solve such PDE models in rather compact Python scripts. We call such a Python script a \emph{FEniCS application code}. In this paper, we develop and document a new software called FEniCS-preCICE adapter, which allows researchers to easily couple FEniCS application codes to other simulation software by using preCICE.

preCICE itself is a library. Thus, for coupling, a participant needs to call the API of preCICE in its source code. Due to the high abstraction level of the preCICE API, the necessary changes to a participant's source code are minimally invasive. These changes are typically realized in a so-called \emph{adapter} -- an additional class, module, or callback library of the participant's source code. An adapter defines the coupling meshes, handles coupling boundary conditions, and realizes the steering of the coupling.
Till around 2016, preCICE users had to write their own adapters for codes they wanted to couple. As many users coupled the same codes, they had to solve the same problems, and continuously reinvented the wheel. Therefore, official, stand-alone, general-purpose adapters for widely used software (e.g.\ for OpenFOAM \cite{weller1998tensorial} or SU2 \cite{economon2016}) were first introduced in \cite{Uekermann2017} to end the waste of precious human development resources. These developments have significantly contributed to the usability and popularity of preCICE and, thus, to the process of scientific discovery.  

In this paper, we follow the same lines of argumentation and introduce a stand-alone, general-purpose adapter for FEniCS. 
We follow a FEniCS-native style, which makes the entry barrier for users of FEniCS as low as possible. The adapter can be easily integrated in existing FEniCS codes. The distributed-memory parallelization of FEniCS is supported out of the box.
FEniCS itself is not a single simulation code, but a library as well. This makes a general-purpose adapter a challenging task. We, therefore, restrict the generality: we focus on 2D surface-coupled problems and study time-dependent fluid-structure interaction (FSI) and conjugate heat transfer (CHT) as examples. The proposed software design, however, is easy to extend, in particular to 3D scenarios and to volume-coupled problems. 

FEniCS has already been used to solve FSI or more general multi-physics problems in a \emph{monolithic} fashion \cite{2019FEniCSMixedDim, FEniCS-HPC, Bergersen2020}. To this end, the library multiphenics\footnote{\url{https://mathlab.sissa.it/multiphenics}} provides FEniCS-tools to solve multi-physics problems. The interaction of the different physical phenomena is modeled through a large coupled equation system that is solved using FEniCS. Very often conforming meshes are required across the different physical domains. We, however, follow a \emph{partitioned} approach, which allows to combine several specialized single-physics participants to solve overall multi-physics problems. FEniCS has already been coupled in a partitioned fashion to reaktoro \cite{Damiani2020} to simulate reactive transport or to other instances of FEniCS to allow for non-conforming meshes in FSI \cite{Massing2015}. In this paper, we present a software for coupling of FEniCS-based applications to arbitrary other simulation software via preCICE.

We describe the software in sufficient detail in Section \ref{sec:software}. This includes an overview of the software architecture, a detailed description of the software's functionality, an exemplary use of its API, and a brief overview of used testing approaches. We give two illustrative use cases of the FEniCS-preCICE adapter in Section \ref{sec:examples}: a conjugate heat transfer simulation with FEniCS and OpenFOAM and a fluid-structure interaction with FEniCS and SU2. Afterwards, we discuss the impact of the new software in Section \ref{sec:impact}.

\section{Software description}
\label{sec:software}
The FEniCS-preCICE adapter might be considered an unusual research software. It is neither a stand-alone single program, nor a library that can be used in such a program. Instead, it is a middle layer software between two large software packages: the finite element library FEniCS and the coupling library preCICE. Furthermore, a coupled simulation, by definition, consists of multiple participants, which brings even more software packages to the table. When describing the FEniCS-preCICE adapter, it is therefore essential to describe how the new software interfaces with these other software packages. In Section \ref{ssec:SoftwareArchitecture}, we give an overview of the overall software architecture alongside a brief introduction to the individual packages: preCICE, FEniCS, and the FEniCS-preCICE adapter. Afterwards, in Section \ref{ssec:SoftwareFunctionalities}, we give detailed information on the API of the FEniCS-preCICE adapter and its functionality. A short example code in Section \ref{ssec:snippets} completes the software description. Last, in Section \ref{ssec:testing}, we explain how the new software is tested.

The software is developed on github\footnote{\url{https://github.com/precice/fenics-adapter}} and the source code is publicly available under LGPL-3.0 license. A Python package is published and maintained on PyPI\footnote{\url{https://pypi.org/project/fenicsprecice}}.

\subsection{Software Architecture}
\label{ssec:SoftwareArchitecture}

\begin{figure}
\center
\tikzstyle{doc}=[%
draw,
thick,
align=center,
color=black,
shape=document,
minimum width=10mm,
minimum height=14.1mm,
shape=document,
inner sep=0ex,
]

\begin{tikzpicture}[scale=1.5]


\coordinate(fenicsOrigin) at (8.5,0);
\draw[fill=ci_pantone283](fenicsOrigin) rectangle node{\mintinline{python}{fenics}} ++(1, 1) [add reference=fenics];

\coordinate(solverOrigin) at ($(fenicsOrigin)-(1.75,0)$);
\draw[fill=ci_pantone300](solverOrigin) rectangle node{\mintinline{bash}{solver.py}} ++(1.5, 1) [add reference=solver];

\coordinate(adapterOrigin) at ($(solverOrigin)-(.25,0)$);
\draw[fill=participantAcolor](adapterOrigin) -- node(adapteraim)[below]{} ++ (-.75,0) -- node(adapterwest)[left]{} ++(0,.27) -- ++(0.375,0) -- ++(0,.46) -- ++(-.375,0) -- ++(0,.27) -- ++(.75,0) [add reference=adapter] -- (adapterOrigin);

\node at([xshift = -1.6cm, yshift=-.3cm]adapterwest) (adapterlabel){\mintinline{bash}{FEniCS-preCICE adapter}};
\draw ($(adapterwest)+(.2, 0)$) -- ($(adapterwest)+(.2, -.3)$) -- (adapterlabel);

\draw[->,yshift=.05cm]([yshift=-.08cm,xshift=+.05cm]solver south) to [out=-30,in=-150] ($([yshift=-.08cm]fenics south)$);
\node at ($0.5*([yshift=.08cm]solver south) + 0.5*([yshift=.08cm]fenics south) + (.4,-.6)$) {\mintinline{python}{import fenics}};

\draw[->,yshift=.05cm]([yshift=.08cm]adapter north) to [out=30,in=150]($([yshift=.08cm]fenics north)$);
\node at ($0.5*([yshift=.08cm]adapter north) + 0.5*([yshift=.08cm]fenics north) + (0,.6)$) {\mintinline{python}{import fenics}};

\coordinate(preciceOrigin) at ($(adapterOrigin)+(-1.5,0)$);
\draw[fill=preciceColor](preciceOrigin) -- ++(.34,0) node[below,pos=.5](preciceaim){} -- ++(0,.27) -- ++(0.375,0) -- ++(0,.46) -- ++(-.375,0) -- ++(0,.27) -- ++(-.34,0) node[above,pos=.5](precicenorth){} -- (preciceOrigin) node[left,pos=.9](precicewest){} node[left,pos=.5](precicemidwest){};

\draw[->,yshift=.05cm]($([yshift=.08cm]adapter north)+(-.1cm, 0)$) to [out=150,in=30] (precicenorth)
node[above left, align=center, xshift=1.2cm, yshift=0.25cm](precicepython){\mintinline{python}{import precice}};

\draw[->] ([yshift=-.08cm,xshift=-.05cm]solver south) to [out=-150,in=-30] (adapteraim)
node[below left, align=right, yshift=-.3cm, xshift=1.5cm](fenicsprecice){
\mintinline{python}{import fenicsprecice}};

\coordinate(otherPreciceOrigin) at ($(preciceOrigin)+(-2.5,0)$);
\draw[fill=preciceColor!50, dashed](otherPreciceOrigin) -- ++(-.34,0) -- ++(0,.27) -- ++(-0.375,0) -- ++(0,.46) -- ++(.375,0) -- ++(0,.27) -- node[pos=.5,above](otherPrecicenorth){} ++(.34,0) -- (otherPreciceOrigin) node[right,pos=.9](otherPreciceeast){} node[right,pos=.5](otherPrecicemideast){};
\node[align=center] at($.5*(precicewest) + .5*(otherPreciceeast)$) (precicelabel){\mintinline{bash}{libprecice}};
\draw ($(precicewest)+(.2,0)$) -- (precicelabel);
\draw[dashed] ($(otherPreciceeast)+(-.2,0)$) -- (precicelabel);
\draw[dashed,<->] (otherPrecicemideast) -- node[above, align=center]{coupling to \\ OpenFOAM, SU2, ...} (precicemidwest);

\node(xmlconfig)[doc] at ($.5*(preciceOrigin) + .5*(otherPreciceOrigin) + (0,2)$){\mintinline{bash}{xml}};
\draw[->] (precicepython.north) -- (precicepython.north |- xmlconfig.east)-- node[above, anchor=south west]{\mintinline{python}{precice.SolverInterface(...)}}(xmlconfig);
\draw[->, dashed] (otherPrecicenorth) -- (otherPrecicenorth.north |- xmlconfig.west) -- (xmlconfig);

\node(jsonconfig)[below=of xmlconfig, doc, yshift=-2cm] {\mintinline{bash}{json}};
\draw[->] (fenicsprecice.south) -- (fenicsprecice.south |- jsonconfig.east)-- node[below, anchor=north west]{\mintinline{python}{fenicsprecice.Adapter(...)}}(jsonconfig);

\end{tikzpicture}
\caption{Overview of software architecture. From left to right: preCICE, FEniCS-preCICE adapter, application code \mintinline{bash}{solver.py}, and FEniCS.}
\label{fig:adapter}
\end{figure}
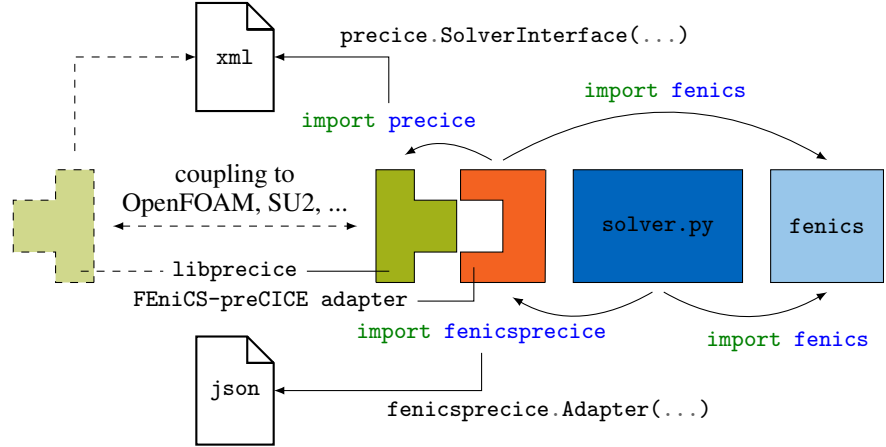

Figure \ref{fig:adapter} gives an overview on how all software layers play together in a coupled simulation using the FEniCS-preCICE adapter.
The user provides a FEniCS application code (\mintinline{bash}{solver.py}), which uses FEniCS (\mintinline{python}{import fenics}) for solving a certain PDE model with a finite element method.
Additionally, the application code imports the FEniCS-preCICE adapter (\mintinline{python}{import fenicsprecice}), which is a Python package itself, for coupling to other simulation software.
The adapter, in turn, imports preCICE (\mintinline{python}{import precice}) -- more specifically the Python bindings of preCICE. Finally, preCICE handles the coupling to other simulation software, for example OpenFOAM or SU2.
Let us have a closer look at the individual packages.\\

\noindent
\emph{preCICE} 
provides three building blocks for coupling mesh-based PDE solvers:
\begin{inparaenum}[1)]
\item Methods for data mapping between non-matching meshes.
\item Fixed-point acceleration methods to stabilize coupled equation systems.
\item Communication between participants, which are separate executables, potentially running on different nodes in a heterogeneous compute cluster. 
\end{inparaenum}
preCICE is a library and follows a peer-to-peer coupling concept -- no central server-like entity is required. The library is written in C++, but also offers bindings in Python generated from the C++ API with Cython \cite{behnel2011cython}.
As input and output arguments, the Python bindings use primitive Python data types as well as NumPy arrays \cite{NumPy}.
preCICE is configured at run time through an \mintinline{bash}{xml} file describing a complete coupled simulation setup\footnote{See \url{https://www.precice.org/configuration-overview.html} for details on configuration of preCICE.}.\\

\noindent
\emph{FEniCS}
is a finite element Python package with an extensive C++ implementation named DOLFIN under the hood. Its API uses a high abstraction level. 
FEniCS provides functionality and data structures for generating and managing meshes. On top of meshes, various finite element spaces can be created. Weak forms of arbitrary PDEs can then be defined in a very compact notation. In Section \ref{ssec:snippets}, we give a brief code example, which illustrates the FEniCS API. The FEniCS-preCICE adapter has been tested and developed using FEniCS 2019.1.0.
For more information on FEniCS, we refer the reader to \cite{LoggMardalEtAl2012}.\\

\noindent
\emph{FEniCS-preCICE}
provides an API for coupling a FEniCS application code using preCICE.
This means, the  FEniCS application code calls the FEniCS-preCICE adapter, which in turn calls preCICE.
Some of the API calls of the FEniCS-preCICE adapter are simply redirected to preCICE. 
For other API calls, the adapter provides substantial functionality, for example for conversion of data structures.
We give a detailed explanation of the API of the adapter in Section \ref{ssec:SoftwareFunctionalities}.
The adapter is configured through a \mintinline{bash}{.json} file, which describes what data is written and read by the adapter.
For simplification, currently only one read and one write data field on a single coupling mesh is technically supported by the adapter, whereas the preCICE API offers more flexibility. An extension in this direction is planned. Furthermore, FEniCS-preCICE was only tested with continuous Galerkin first and second order finite elements, while FEniCS provides a broad range of available finite element spaces, including higher order and discontinuous Galerkin.

\subsection{Software Functionalities}
\label{ssec:SoftwareFunctionalities}

The FEniCS-preCICE adapter uses a central \mintinline{python}{Adapter} object as handle to its API. We list the important parts of the API in Listing \ref{listing:API} and describe individual member functions of \mintinline{python}{Adapter} step by step in the following. To follow all implementation details of this section requires some previous knowledge of FEniCS.  

\begin{listing}
\begin{minted}[escapeinside=!!]{python}
import precice  # python bindings of preCICE

class Adapter:
    def __init__(self, adapter_config_filename='precice-adapter-config.json'):!\label{adapter_init_start}!
        # create adapter object and interface to preCICE python bindings
        self._interface = precice.Interface(...)

    def initialize(self, coupling_subdomain, read_function_space=None, write_object=None):
        # initialize coupling mesh and initialize data in preCICE
        precice_dt = self._interface.initialize()
        ...
        self._interface.initialize_data()
        return precice_dt  # returns maximum allowed timestep size!\label{adapter_init_end}!

    def read_data(self):!\label{data_start}!
        # creates a dictionary with coupling vertices and reads associated data from preCICE
        return data

    def write_data(self, write_function):
        # sample write_function and write data to preCICE !\label{data_end}!

    def create_coupling_expression(self):!\label{consistent_start}!
        # create uninitialized CouplingExpression
        return CouplingExpression(...)

    def update_coupling_expression(self, coupling_expression, data):
        # get nodal data and vertex coordinates from dictionary data 
        coupling_expression.update_boundary_data(nodal_data, x_coordinates, y_coordinates)!\label{consistent_end}!

    def get_point_sources(self, data):!\label{conservative_start}!
        # creates one list of PointSources per dimensions from given coupling data
        return x_PointSources, y_PointSources !\label{conservative_end}!

    def store_checkpoint(self, user_u, t, n):!\label{checkpointing_start}!
        self._checkpoint = SolverState(user_u.copy(), t, n)
        self._interface.mark_action_fulfilled(precice.action_write_iteration_checkpoint())

    def retrieve_checkpoint(self):
        self._interface.mark_action_fulfilled(precice.action_read_iteration_checkpoint())
        return self._checkpoint.get_state() !\label{checkpointing_end}!

    def advance(self, dt):!\label{steering_start}!
        return self._interface.advance(dt)!\label{steering_end}!
\end{minted}
\caption{API of the FEniCS-preCICE adapter (excerpt). The actual implementation is sketched, whenever it is short enough.}
\label{listing:API}
\end{listing}

\paragraph{Adapter initialization (lines \ref{adapter_init_start} -- \ref{adapter_init_end})}

The constructor of \mintinline{python}{Adapter} creates and configures the adapter object. The object is initialized by calling \mintinline{python}{initialize}, where the argument \mintinline{python}{coupling_subdomain} is the domain boundary where data should be coupled. Two optional arguments exist to configure the coupling: For one-way coupling either a \mintinline{python}{read_function_space} or a \mintinline{python}{write_object} must be provided - depending on whether the participant reads or writes data. For two-way coupling both arguments are required. 

A FEniCS \mintinline{python}{FunctionSpace} is provided for the \mintinline{python}{read_function_space} and the \mintinline{python}{write_object} to provide the coupling mesh as well as the type of function (scalar/vector, order) to the adapter. If the user wants to provide initial write data, a FEniCS \mintinline{python}{Function} can be provided as \mintinline{python}{write_object}.

\paragraph{Data access (line \ref{data_start} -- \ref{data_end})}

The user can call \mintinline{python}{write_data} and \mintinline{python}{read_data} to write data to or read data from preCICE, respectively. 
The argument \mintinline{python}{write_function} provided to \mintinline{python}{write_data} is a FEniCS \mintinline{python}{Function}, which the adapter samples at the coupling mesh. Calling \mintinline{python}{read_data} returns a dict \mintinline{python}{data} that contains coupling mesh points and associated data. It can be used to directly update coupling expressions or point sources.

\paragraph{Coupling consistent quantities via expressions (lines \ref{consistent_start} -- \ref{consistent_end})}

To enforce coupling boundary conditions, we need to distinguish two types of coupling data: quantities that require consistent data mapping (e.g.\ temperature, heat flux, or force per unit volume) and quantities that require conservative data mapping (e.g.\ forces). Let us first consider boundary conditions for consistent quantities. 
For this case, FEniCS' \mintinline{python}{Expression} has proven to be the right tool. This object is very flexible and can be used in many different ways -- not only for boundary conditions. The FEniCS book \cite{LoggMardalEtAl2012} gives many examples how to use an expression for both, Dirichlet and Neumann boundary conditions. An essential Dirichlet boundary condition, for instance, can be defined by FEniCS' \mintinline{python}{DirichletBC}. However, expressions can also be used directly in the weak form, for instance for Neumann boundary conditions (\mintinline{python}{... + expr * v * ds}) or for volume terms (\mintinline{python}{... + expr * v * dx}). Here, \mintinline{python}{expr} is an \mintinline{python}{Expression}, \mintinline{python}{v} the test function, and \mintinline{python}{ds} and \mintinline{python}{dx} a surface and volume integration element. 
Normally, for boundary conditions, expressions are explicitly given as symbolic expressions (e.g.\ \mintinline{python}{Expression('sin(x[0]) + cos(x[1])')}). For coupling boundaries, however, such a continuous representations needs to be constructed from the nodal values given by preCICE. To this end, the adapter provides a \mintinline{python}{CouplingExpression} which inherits from FEniCS' \mintinline{python}{UserExpression}. The continuous representation is constructed using an interpolation routine following the approach suggested in \cite{Lindner2017RBF}, where first a polynomial least-squares fit is constructed followed by a radial-basis function interpolation. For the radial-basis function interpolation, we use routines from SciPy \cite{SciPy}. This interpolation is not to be confused with the data mapping that preCICE uses. An uninitialized \mintinline{python}{CouplingExpression} is created and returned by \mintinline{python}{create_coupling_expression}. It is initialized or updated by calling \mintinline{python}{update_coupling_expression} and providing the dict \mintinline{python}{data} obtained from \mintinline{python}{read_data}.

\paragraph{Coupling conservative quantities via point sources (lines \ref{conservative_start} -- \ref{conservative_end})}

Boundary conditions for conservative quantities can be realized by point-wise boundary conditions \cite{Hertrich2019}. To this end, FEniCS offers \mintinline{python}{PointSource(V, p, magnitude)}, which can be applied to the equation system by \mintinline{python}{PointSource.apply(rhs)}. Here, \mintinline{python}{V} is a function space, \mintinline{python}{p} the coordinates where the point source is applied, and \mintinline{python}{magnitude} the magnitude of the point source.
\mintinline{python}{get_point_sources} allows the user to obtain point sources at the individual coupling mesh points; again by providing the dict \mintinline{python}{data} obtained from \mintinline{python}{read_data}.
Please note that, contrary to a \mintinline{python}{CouplingExpression}, which is created once and then updated via a pointer-like access pattern, a \mintinline{python}{PointSource} is just overwritten. 

\paragraph{Checkpointing (lines \ref{checkpointing_start} -- \ref{checkpointing_end})}

For implicit coupling (also known as strong or tight coupling), each time step (or even multiple time steps within a so-called time window) needs to be repeated iteratively until the coupling residual drops below a defined threshold. preCICE handles the convergence measurement, the iteration control, and the acceleration of the implicit coupling loop.
The responsibility of the adapter is to provide a mechanism to go backwards in time. This is realized by storing and retrieving checkpoints of the complete solver state. To this end, the adapter offers methods \mintinline{python}{store_checkpoint} and \mintinline{python}{retrieve_checkpoint}, which are designed such that a user cannot accidentally destroy or overwrite checkpoints.

\paragraph{Steering (lines \ref{steering_start} -- \ref{steering_end})}
Steering methods, which allow to control the time and the coupling loop (\mintinline{python}{advance} and several others), are all directly forwarded to preCICE. We refer to the documentation of the Python bindings of preCICE \footnote{\url{https://github.com/precice/python-bindings}} for details.

\paragraph{Parallelization}
FEniCS supports distributed-memory parallelization based on MPI:
\begin{minted}[linenos=off]{bash}
    mpiexec -np numberOfRanks python3 solver.py
\end{minted}
The FEniCS-preCICE adapter directly supports this parallelization. We want to note a few necessary implementation details.
In most cases, the domain decomposition of FEniCS does not yield a situation where all parallel ranks are located at the coupling boundary. Nevertheless to allow for a single implementation for serial and parallel cases, the adapter object is created and initialized on all ranks. If the domain of a rank is not connected to the coupling boundary, the rank is considered as an inactive rank from the perspective of the adapter.
A further technical challenge results from the ghost communication layers in FEniCS. As standard for the finite element method, vertices at the domain boundaries are duplicated on all connected ranks. However, only one rank has the ownership of a specific vertex. At the coupling boundary, only the rank which owns a vertex defines it as part of the coupling mesh for preCICE. Thus, in the adapter, each rank can only read new values from preCICE on the vertices it owns. The reconstruction of the coupling boundary condition, however, requires values also at non-owned vertices. To exchange these values between ranks, the adapter uses its own communication step after reading values from preCICE.
The MPI wrapper shipped with FEniCS cannot be used here because the communication needs to take place after data is read from preCICE and before it is updated in FEniCS.
The current parallel implementation, however, only supports boundary conditions defined in the form of expressions. Generating point source objects in parallel cannot be supported due to a known problem in FEniCS.

\subsection{Example: A Simple Coupled Heat Transfer Solver}
\label{ssec:snippets}

To complete the explanation of the software architecture and the API in the last sections, we now show a simple, but almost complete example application code \mintinline{bash}{solver.py}. The example we study is the time-dependent heat equation and is borrowed from the FEniCS tutorials \cite{Langtangen2016}.

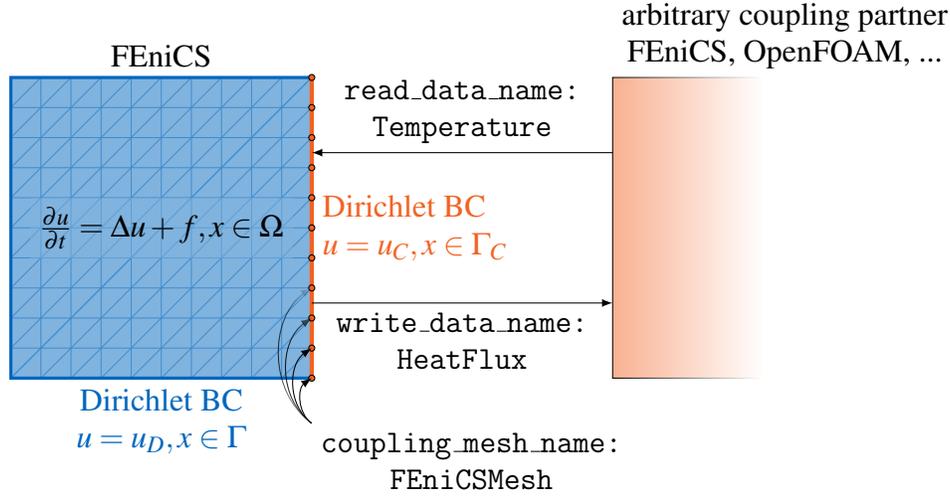
\begin{figure}
\begin{center}
\begin{tikzpicture}[scale=4, every node/.style={font=\large}]

\coordinate(origin) at (0,0);

\coordinate(bottomRight) at (1,0);
\coordinate(topLeft) at (0,1);
\coordinate(topRight) at (1,1);

\coordinate(partnerOrigin) at (2,0);
\coordinate(partnerTopRight) at (2.5, 1);
\coordinate(partnerTopLeft) at (2, 1);

\draw[fill=participantBcolor!50] (origin) rectangle (topRight);
\draw[step=0.1,very thin,participantBcolor!75] (origin) grid (topRight);
\foreach \i in {0,...,10}{%
\draw[very thin, participantBcolor!75] ($\i*0.1*(origin)+10*0.1*(bottomRight)-\i*0.1*(bottomRight)$) -- ($\i*0.1*(topRight)+10*0.1*(bottomRight)-\i*0.1*(bottomRight)$);
\draw[very thin, participantBcolor!75] ($\i*0.1*(origin)+10*0.1*(topLeft)-\i*0.1*(topLeft)$) -- ($\i*0.1*(topRight)+10*0.1*(topLeft)-\i*0.1*(topLeft)$);
}
\draw[fill=participantAcolor!50,path fading=fade right] (partnerOrigin) rectangle (partnerTopRight);
\draw[draw=none] (partnerTopLeft) node[above,black,align=center,anchor=south west]{arbitrary coupling partner\\ FEniCS, OpenFOAM, ...} -- (partnerTopRight);
\draw[ultra thick, participantAcolor] (bottomRight) -- node[right,align=left]{Dirichlet BC \\ $u = u_C, x \in \Gamma_C$} (topRight);
\draw[very thick, participantBcolor] (bottomRight) -- node[below,align = center]{Dirichlet BC \\ $u=u_D, x \in \Gamma$} (origin) -- (topLeft) -- node[above, black]{FEniCS} (topRight);
\draw[fill=none,draw=none] (origin) rectangle node[fill opacity=0.5,fill=participantBcolor!50,text opacity=1]{$\frac{\partial u}{\partial t} = \Delta u + f, x\in \Omega$} (topRight);

\foreach \i in {0,...,10}{%
\draw[fill=participantAcolor] ($\i*0.1*(bottomRight)+10*0.1*(topRight)-\i*0.1*(topRight)$) circle [radius=0.01];
}

\draw[<-,black] ($(bottomRight) + (0,.1)$) to[in=145, out=-145] (1,-.15);
\draw[<-,black] ($(bottomRight) + (0,0)$) to[in=145, out=-145] (1,-.15) node[align=center, right,anchor=north west]{\texttt{coupling\_mesh\_name:}\\ \texttt{FEniCSMesh}};
\draw[<-,black,path fading=fade middle] ($(bottomRight) + (0,.2)$) to[in=145, out=-145] (1,-.15);
\draw[<-,black,path fading=fade right] ($(bottomRight) + (0,.3)$) to[in=145, out=-145] (1,-.15);

\draw[->] (1,.25) -- node[align=center, below]{\texttt{write\_data\_name:}\\ \texttt{HeatFlux}} (2,.25);
\draw[<-] (1,.75) -- node[align=center, above]{\texttt{read\_data\_name:}\\ \texttt{Temperature}} (2,.75);

\end{tikzpicture}
\end{center}
\caption{Unit square $\Omega$, where heat equation is solved by the FEniCS participant. Dirichlet boundary conditions are applied on $\Gamma$. Coupling happens on $\Gamma_C$, where another Dirichlet boundary condition is applied and heat flux is sampled to establish the coupling.}
\label{fig:heat_setup}
\end{figure}

We define one edge of the squared domain as a Dirichlet coupling boundary. Through this boundary, our example code could be coupled to another FEniCS application code solving the time-dependent heat equation as well, which would render the overall problem into a simple partitioned heat equation. Alternatively, the coupling partner could also be a fluid solver, which would result in a conjugate heat transfer scenario. We show the results of such an example in Section \ref{cht}. The setup is illustrated in \autoref{fig:heat_setup}. In the following, we first look briefly at the adapter configuration and, afterwards, we discuss the application code.

\subsubsection{Configuration of the Adapter}

Listing \ref{listing:adapter_config} shows the configuration file of our example. The coupling participant is named \mintinline{bash}{FEniCS} and the associated coupling mesh \mintinline{bash}{FEniCSMesh}. 
The participant reads temperature values from preCICE and uses them to construct a Dirichlet boundary condition at the coupling boundary. Finally, the participant writes heat flux values to preCICE, which are explicitly computed as gradients of the temperature field.

\begin{listing}[h!]
\begin{center}
\begin{minipage}{0.5\textwidth} 
\begin{minted}{json}
{
"participant_name": "FEniCS",
"config_file_name": "precice-config.xml",
"interface": {
    "coupling_mesh_name": "FEniCSMesh",
    "write_data_name": "HeatFlux",
    "read_data_name": "Temperature"
  }
}
\end{minted}
\end{minipage}
\end{center}
\caption{Adapter configuration file \mintinline{bash}{precice-adapter-config.json}}
\label{listing:adapter_config}
\end{listing}

\subsubsection{Application Code}

\begin{listing}
\hfill 
\begin{minipage}{\textwidth} 
\begin{minted}[escapeinside=!!, 
               mathescape=true,
               highlightlines={\getrefnumber{import_fenicsprecice}-\getrefnumber{import_fenicsprecice}, 
                               \getrefnumber{coupling_boundary}-\getrefnumber{coupling_boundary},
                               \getrefnumber{init_adapter_start}-\getrefnumber{init_adapter_end},
                               \getrefnumber{coupling_expression_start}-\getrefnumber{coupling_expression_end},
                               \getrefnumber{assign_dt}-\getrefnumber{assign_dt},
                               \getrefnumber{time_loop}-\getrefnumber{time_loop},
                               \getrefnumber{write_cp_start}-\getrefnumber{write_cp_end},
                               \getrefnumber{read_start}-\getrefnumber{read_end}, 
                               \getrefnumber{write_start}-\getrefnumber{write_end}, 
                               \getrefnumber{read_cp_start}-\getrefnumber{read_cp_end},
                               \getrefnumber{advance}-\getrefnumber{advance},
                               \getrefnumber{function_space_flux}-\getrefnumber{function_space_flux}}]{python}
from fenics import *
from fenicsprecice import Adapter                         !\label{import_fenicsprecice}!
import numpy as np

mesh = UnitSquareMesh(10, 10)   !\label{geo_start}!
class Boundary(SubDomain): ...   !\label{geo_end}!
!\tikzmark{cpl_mesh}!class CouplingBoundary(SubDomain): ... !\label{coupling_boundary}!

V = V_bc = FunctionSpace(mesh, 'P', 2)  !\label{function_space_start}!
u, v = TrialFunction(V), TestFunction(V) !\label{function_space_end}!
V_flux = VectorFunctionSpace(mesh, 'P', 1) !\label{function_space_flux}!
u_D = Expression('...',degree=2)  !\label{DirBC_start}!# in our example constant over time
uncoupled_bc = DirichletBC(V_bc, u_D, Boundary)  !\label{DirBC_end}!

!\tikzmark{init_start}!adapter = Adapter("precice-adapter-config.json")  !\label{init_adapter_start}!
precice_dt = adapter.initialize(CouplingBoundary, read_function_space=V_bc, write_object=V_flux) !\label{init_adapter_end}!
!\tikzmark{cpl_read}!u_C = adapter.create_coupling_expression() !\label{coupling_expression_start}!
!\tikzmark{init_end}!coupled_bc = DirichletBC(V_bc, u_C, CouplingBoundary) !\label{coupling_expression_end}! 

fenics_dt = 0.1  # time step size required by FEniCS
dt = Constant(0)  # time step size used by this participant 
u_n = interpolate(u_D, V)  # define initial value for $u^n$
u_np1 = Function(V)  # define function for solution $u^{n+1}$
f = Expression(...)  # right-hand side, in our example constant over time
F = u*v*dx + dt*dot(grad(u), grad(v))*dx - (u_n + dt*f)*v*dx  !\label{weak_form}!
t = 0  # initialize time

while adapter.is_coupling_ongoing():                                 !\label{time_loop}!
!\tikzmark{write_cp_start}!  if adapter.is_action_required(adapter.action_write_iteration_checkpoint()):!\label{write_cp_start}!
!\tikzmark{write_cp_end}!    adapter.store_checkpoint(solution=u_n, t=t)                      !\label{write_cp_end}!

!\tikzmark{cpl_start}!!\tikzmark{read_start}!  read_data = adapter.read_data()                                    !\label{read_start}!
!\tikzmark{read_end}!  adapter.update_coupling_expression(u_C, read_data)!\label{read_end}!
  dt.assign(np.min([fenics_dt, precice_dt]))                         !\label{assign_dt}!
  solve(lhs(F) == rhs(F), u_np1, [uncoupled_bc, coupled_bc])         !\label{solve}!
!\tikzmark{cpl_write}!!\tikzmark{write_start}!  flux = some_postprocessing(u_np1, V_flux)!\label{write_start}!
!\tikzmark{write_end}!  adapter.write_data(flux)!\label{write_end}!
!\tikzmark{cpl_end}!  precice_dt = adapter.advance(dt(0))                                !\label{advance}!

!\tikzmark{read_cp_start}!  if adapter.is_action_required(adapter.action_read_iteration_checkpoint()): !\label{read_cp_start}!
    u_cp, t_cp = adapter.retrieve_checkpoint()
    u_n.assign(u_cp)
!\tikzmark{read_cp_end}!    t = t_cp                                                         !\label{read_cp_end}!
  else:
    u_n.assign(u_np1)
    t += float(dt)
\end{minted}
\end{minipage}
\caption{Coupled heat equation FEniCS application code. The calls to FEniCS-preCICE are highlighted, while the remaining lines represent a simplified version of the original heat equation example from the FEniCS tutorials \cite{Langtangen2016}.
}
\label{listing:CoupledHeatEquation}
\begin{scriptsize}
\begin{tikzpicture}[
remember picture,
B/.style args = {#1/#2}{
            decorate,
            decoration={calligraphic brace, amplitude=5pt,
                        pre=moveto,pre length=1pt,post=moveto,post length=1pt,
                        raise=#1,
                        #2,
                        },
            thick}]
\draw[B=1mm/mirror ,overlay] ([yshift=.8em,xshift=-.6cm]pic cs:init_start) -- ([yshift=-.4em,xshift=-.6cm]pic cs:init_end) node[midway, rotate=90,yshift=.5cm+.5em,align=center] {initialize adapter \& \\ coupling boundary condition};
\draw[B=1mm/mirror ,overlay] ([yshift=.8em,xshift=-.6cm]pic cs:write_cp_start) -- ([yshift=-.4em,xshift=-.6cm]pic cs:write_cp_end) node[midway, rotate=90,yshift=.5cm+.5em,align=center,xshift=1em] {write\\ checkpoint};
\draw[B=1mm/mirror ,overlay] ([yshift=.8em,xshift=-.6cm]pic cs:write_start) -- ([yshift=-.4em,xshift=-.6cm]pic cs:write_end) node[midway, rotate=90,yshift=.5cm+.5em,align=center] {write\\ data};
\draw[B=1mm/mirror ,overlay] ([yshift=.8em,xshift=-.6cm]pic cs:read_start) -- ([yshift=-.4em,xshift=-.6cm]pic cs:read_end) node[midway, rotate=90,yshift=.5cm+.5em,align=center] {read\\ data};
\draw[B=1mm/mirror ,overlay] ([yshift=.8em,xshift=-.6cm]pic cs:read_cp_start) -- ([yshift=-.4em,xshift=-.6cm]pic cs:read_cp_end) node[midway, rotate=90,yshift=.5cm+.5em,align=center] {read\\ checkpoint};
\end{tikzpicture}
\end{scriptsize}
\end{listing}

Listing \ref{listing:CoupledHeatEquation} shows the almost complete application code of our example. Let us first have a look at the FEniCS code parts. Lines \ref{geo_start} and \ref{geo_end} define the geometry and mesh. Afterwards, lines \ref{function_space_start} and \ref{function_space_end} define the (quadratic) function space as well as trial and test functions, which are then used in line \ref{weak_form} to define the weak form of the time-dependent heat equation. For time integration, an implicit Euler scheme is used.
Furthermore, lines \ref{DirBC_start} and \ref{DirBC_end} define Dirichlet boundary conditions at the non-coupling boundary. The discretized linear system is solved in every timestep in line \ref{solve}.

The highlighted lines show the code related to the coupling. We import FEniCS-preCICE in line \ref{import_fenicsprecice}. Line \ref{coupling_boundary} defines a subdomain as coupling boundary. Then, lines \ref{init_adapter_start} to \ref{coupling_expression_end} create the handle to the adapter, initialize it with a read and write function space used for two-way coupling, and create the Dirichlet coupling boundary condition. In line \ref{time_loop}, the time loop control is handed over to preCICE. If required, data checkpoints are written in lines \ref{write_cp_start} and \ref{write_cp_end} and read in lines \ref{read_cp_start} to \ref{read_cp_end}. Temperature values are read from preCICE in line \ref{read_start} and used to update the coupling boundary condition in line \ref{read_end}.
After solving the linear system, heat fluxes are extracted from the new solution (using the \mintinline{python}{VectorFunctionSpace} from line \ref{function_space_flux}) and written to preCICE in lines \ref{write_start} and \ref{write_end}.
Then, finally the actual coupling is advanced. Here, preCICE returns an upper limit for the next timestep size, which is enforced in line \ref{assign_dt}.

\subsection{Testing}\label{ssec:testing}

The complex software architecture of the FEniCS-preCICE adapter as introduced in Section \ref{ssec:SoftwareArchitecture} makes testing the software more involved than testing most other research software. First, the FEniCS-preCICE adapter is a middle software layer -- it is called by a FEniCS application code and calls in turn preCICE (via its Python bindings). Second, a coupled simulation needs at least two participants, thus at least two layered software stacks, of which the FEniCS-preCICE adapter could be part of one or both. Both concepts are not unusual in software engineering in general. Thus, there are known techniques for testing, such as \emph{mocking} \cite{fowler2007mocks}. The concepts are, however, uncommon for research codes, which typically only consist of stand-alone executables or libraries.
We apply two different approaches to test the FEniCS-preCICE adapter. 

The first approach tests the software in a complete coupled setup. We couple two FEniCS application codes to keep the number of dependencies small. We take the example we just introduced in the previous section and couple it with its counterpart -- a time-dependent heat equation with a Neumann coupling boundary condition. We could then compare the results of this partitioned heat equation with those of a single-domain heat equation in FEniCS. We follow, however, an even simpler approach (similar to the FEniCS tutorials \cite{Langtangen2016}): We construct a manufactured solution with linear dependency in time and quadratic dependency in space. As we use $P^2$ finite elements and first-order time integration, the discretization can recover the exact solution for arbitrary (coarse) space and time resolutions\footnote{Refer to \cite[Section 4.1]{QNWI} for details on the discretization.}. The only remaining error component is the coupling error, which can be controlled by the coupling convergence measures in every timestep. Tightening the thresholds reduces the error until machine precision\footnote{Case is provided under \url{https://github.com/precice/tutorials/tree/a166efa/partitioned-heat-conduction}.}. In preCICE nomenclature, we refer to such a test with multiple participants and preCICE as a dependency as a system test. 

The second approach to test the FEniCS-preCICE adapter regards the software as an isolated unit -- independent of any other coupled participant and even independent of preCICE itself. 
To this end, we use a \emph{mocked up} version of the Python bindings of preCICE. This allows us to test individual functions of the adapter, which then call the mocked up dummy implementation instead of preCICE itself. To illustrate the concept, Listing \ref{lst:testing} gives an example on how to test the adapter function \mintinline{python}{read_data}.  
Line \ref{import_mock} imports the mock object \mintinline{python}{MockedPrecice}.
In line \ref{patch}, we use \mintinline{python}{patch} from \mintinline{python}{unittest.mock}\footnote{\url{https://docs.python.org/3/library/unittest.mock.html}} to replace the real implementation of preCICE with the mock object. This means that in line \ref{import_mock_precice} and indirectly in line \ref{import_mock_adapter}, not the real preCICE, but the mock object is imported. The test can be run without preCICE being installed on the test system.   
In line \ref{mocked_read_data}, we use \mintinline{python}{MagicMock} to define the behavior of the mocked preCICE function \mintinline{python}{read_block_scalar_data}.
We then can test whether \mintinline{python}{read_data} converts to and returns the correct data in line \ref{test_output}. 
\mintinline{python}{MagicMock} also allows us to record the arguments a mocked function receives. We, finally, use this functionality in line \ref{test_input} to also test whether preCICE receives the correct input arguments from the adapter.

\begin{listing}[h!]
\begin{minted}[escapeinside=!!, fontsize=\scriptsize]{python}
from unittest import TestCase
import tests.MockedPrecice !\label{import_mock}!

@patch.dict('sys.modules', **{'precice': tests.MockedPrecice}) !\label{patch}!
class TestReadData(TestCase):
  def test_scalar_read(self):
    from unittest.mock import MagicMock
    from precice import Interface !\label{import_mock_precice}!
    import fenicsprecice !\label{import_mock_adapter}!

    # mock preCICE API
    dummy_data = np.arange(...)  # hard-coded dummy values
    Interface.read_block_scalar_data = MagicMock(return_value=dummy_data) !\label{mocked_read_data}!

    # initialize adapter
    adapter = fenicsprecice.Adapter(self.dummy_config)
    [...]
    # call adapter API
    read_data = adapter.read_data()

    # (1) check whether expected output is received
    assert(read_data == expected_read_data) !\label{test_output}!

    # (2) check whether preCICE API was called with expected arguments
    recorded_args = Interface.read_block_scalar_data.call_args[0]
    expected_args = ...  # hard-coded
    for arg, expected_arg in zip(recorded_args, expected_args):
      assert(arg == expected_arg) !\label{test_input}!
\end{minted}
\caption{Testing the individual function \mintinline{python}{read_data} of the FEniCS-preCICE adapter with a mocked preCICE implementation.}
\label{lst:testing}
\end{listing}

\section{Illustrative Examples}
\label{sec:examples}

We give two examples to showcase how the FEniCS-preCICE adapter can be used in practice. 
First, in Section \ref{cht}, a heat conduction solver in FEniCS is coupled to an OpenFOAM fluid solver for conjugate heat transfer (CHT).
Afterwards, in Section \ref{fsi}, a linear elasticity solver in FEniCS is coupled to an SU2 fluid solver for fluid-structure interaction (FSI). We deliberately pick two different applications of FEniCS (heat transfer and structural mechanics) and two different coupling partners (OpenFOAM, SU2) to show the wide applicability of the new adapter. For both setups, we use simple geometries. These geometrical setups are offered as preCICE tutorials for many different participant combinations (FEniCS, OpenFOAM, SU2, deal.II, Nutils, Code\_Aster, CalculiX). FEniCS-preCICE adapter release v1.0.1\footnote{\url{https://github.com/precice/fenics-adapter/releases/tag/v1.0.1}} is used in these examples.

\subsection{Conjugate heat transfer with FEniCS and OpenFOAM}\label{cht}

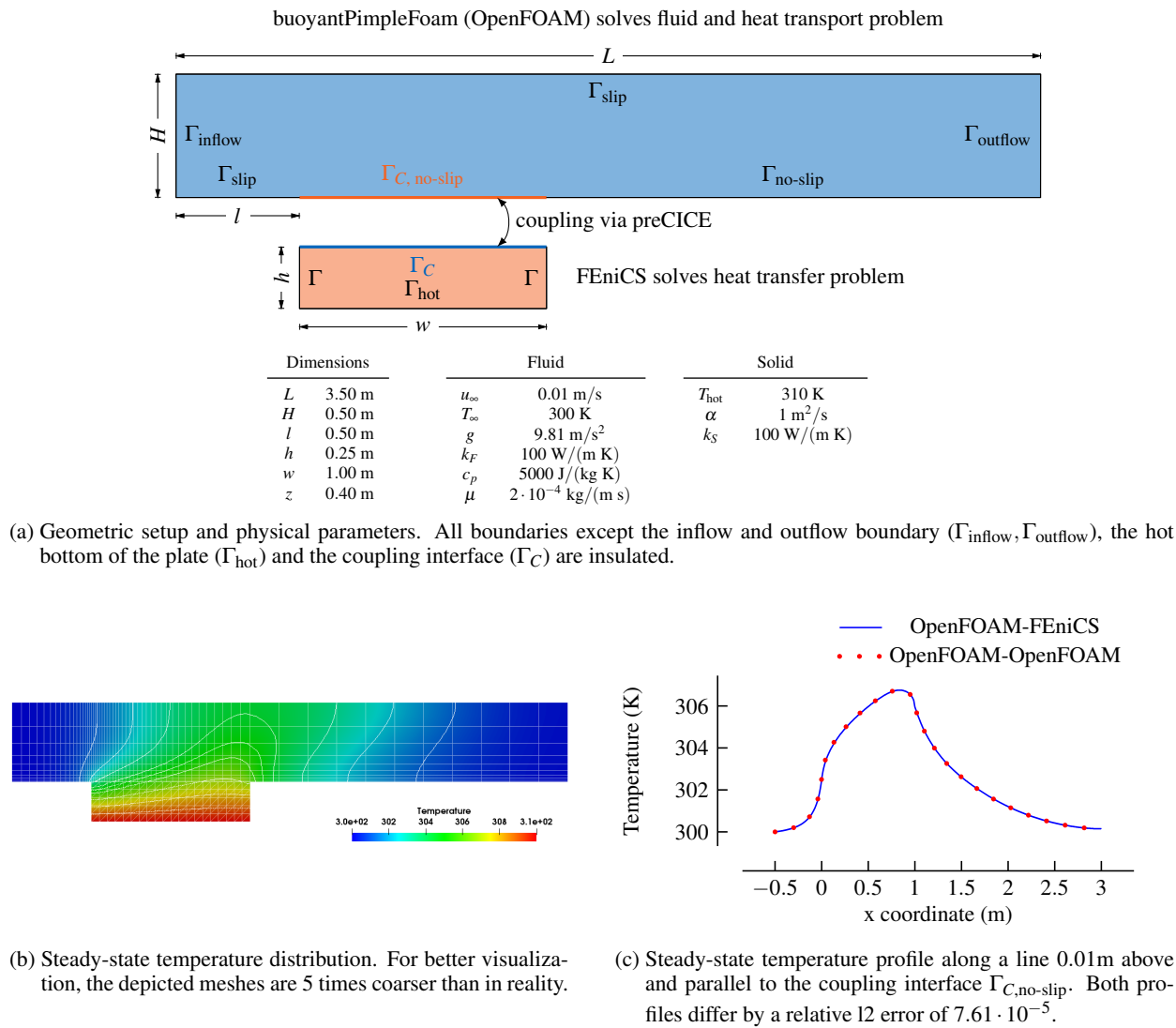
\begin{figure}
\begin{subfigure}{\textwidth}
\begin{minipage}[t]{\textwidth}
\begin{center}
\begin{tikzpicture}[scale=3.5, every node/.style={font=\small}]

\coordinate(origin) at (0,0);

\coordinate(lengthPlate) at (1,0);
\coordinate(thicknessPlate) at (0,-0.25);

\coordinate(leftTopPlate) at ($(origin) + (0,-.2)$);
\coordinate(rightTopPlate) at ($(leftTopPlate)+(lengthPlate)$);
\coordinate(leftBottomPlate) at ($(leftTopPlate)+(thicknessPlate)$);
\coordinate(rightBottomPlate) at ($(rightTopPlate)+(thicknessPlate)$);

\coordinate(channelWidth) at (0,0.5);
\coordinate(channelPrePlate) at (-0.5,0);
\coordinate(channelPostPlate) at (2,0);

\coordinate(leftBottomInflow) at ($(origin)+(channelPrePlate)$);
\coordinate(leftTopInflow) at ($(leftBottomInflow)+(channelWidth)$);

\coordinate(rightBottomOutflow) at ($(origin)+(lengthPlate)+(channelPostPlate)$);
\coordinate(rightTopOutflow) at ($(rightBottomOutflow)+(channelWidth)$);

\draw[fill=participantBcolor!50](leftBottomInflow) rectangle (rightTopOutflow);
\coordinate(midChannel) at ($.5*(leftTopInflow)+.5*(rightTopOutflow)$);
\node[above = .5cm of midChannel]{buoyantPimpleFoam (OpenFOAM) solves fluid and heat transport problem};

\draw[fill=participantAcolor!50](leftTopPlate) rectangle (rightBottomPlate);
\coordinate(midPlate) at ($.5*(rightBottomPlate)+.5*(rightTopPlate)$);
\node[right = .3cm of midPlate, anchor=west]{FEniCS solves heat transfer problem};

\draw[very thick, participantBcolor](leftTopPlate) --  node[below, black, align=center]{\textcolor{participantBcolor}{$\Gamma_C$}} (rightTopPlate);  
\draw[very thick, participantAcolor](origin) -- node[above, black, align=center]{\textcolor{participantAcolor}{$\Gamma_{C\text{, no-slip}}$}} ($(origin)+(lengthPlate)$);  
\draw[<->] ($(origin)+(lengthPlate)+(-.2,0)$) to[in=45, out=-45] node[right]{coupling via preCICE}($(rightTopPlate)+(-.2,0)$);
\draw(leftBottomPlate) -- node[above, black]{$\Gamma_\text{hot}$} (rightBottomPlate);  
\draw(leftTopPlate) -- node[right, black]{$\Gamma$} (leftBottomPlate);  
\draw(rightTopPlate) -- node[left, black]{$\Gamma$} (rightBottomPlate);  

\draw(leftBottomInflow) -- node[above, black, align=center]{$\Gamma_\text{slip}$} (origin);  
\draw(rightBottomOutflow) -- node[above, black, align=center]{$\Gamma_\text{no-slip}$} ($(origin)+(lengthPlate)$);  
\draw(leftTopInflow) -- node[below, black, align=center]{$\Gamma_\text{slip}$} (rightTopOutflow);  
\draw(leftTopInflow) -- node[right, black, align=right]{$\Gamma_\text{inflow}$} (leftBottomInflow);  
\draw(rightBottomOutflow) -- node[left, black, align=left]{$\Gamma_\text{outflow}$} (rightTopOutflow);  

\dimline[extension start length=0, extension end length=0] {($(leftBottomInflow)+(-.21em,0)$)}{($(leftTopInflow)+(-.21em,0)$)}{$H$};
\dimline[extension start length=0, extension end length=0] {($(leftTopInflow)+(0,.21em)$)}{($(rightTopOutflow)+(0,.21em)$)}{$L$};
\dimline[extension start length=0, extension end length=0] {($(leftBottomInflow)-(0,.21em)$)}{($(origin)-(0,.21em)$)}{$l$};

\dimline[extension start length=0, extension end length=0] {($(leftBottomPlate)-(.21em,0)$)}{($(leftTopPlate)-(.21em,0)$)}{$h$};
\dimline[extension start length=0, extension end length=0] {($(leftBottomPlate)-(0,.21em)$)}{($(rightBottomPlate)-(0,.21em)$)}{$w$};

\end{tikzpicture}
\end{center}
\end{minipage}
\begin{minipage}[t]{\textwidth}
\begin{center}
\begin{scriptsize}
\begin{minipage}[t]{.15\textwidth}
\begin{tabular}[t]{cc}
\multicolumn{2}{c}{Dimensions}\\
\midrule
$L$ & $3.50\ \text{m}$\\
$H$ & $0.50\ \text{m}$\\
$l$ & $0.50\ \text{m}$\\
$h$ & $0.25\ \text{m}$\\
$w$ & $1.00\ \text{m}$\\
$z$ & $0.40\ \text{m}$\\
\end{tabular}
\end{minipage}
\begin{minipage}[t]{.2\textwidth}
\begin{tabular}[t]{cc}
\multicolumn{2}{c}{Fluid}\\
\midrule
$u_\infty$ & $0.01\ \text{m}/\text{s}$\\
$T_\infty$ & $300\ \text{K}$\\
$g$ & $9.81\ \text{m}/\text{s}^2$\\  
$k_F$ & $100\ \text{W}/(\text{m K})$\\ 
$c_p$ & $5000\ \text{J}/(\text{kg K})$\\  
$\mu$ & $2\cdot 10^{-4}\ \text{kg}/(\text{m s})$\\  
\end{tabular}
\end{minipage}
\begin{minipage}[t]{.2\textwidth}
\begin{tabular}[t]{cc}
\multicolumn{2}{c}{Solid}\\
\midrule
$T_\text{hot}$ & $310\ \text{K}$\\
$\alpha$ & $1\ \text{m}^2/\text{s}$ \\  
$k_S$ & $100\ \text{W}/(\text{m K})$ \\  
\end{tabular}
\end{minipage}
\end{scriptsize}
\end{center}
\end{minipage}
    \caption{Geometric setup and physical parameters. All boundaries except the inflow and outflow boundary ($\Gamma_\text{inflow}, \Gamma_\text{outflow}$), the hot bottom of the plate ($\Gamma_\text{hot}$) and the coupling interface ($\Gamma_C$) are insulated.}
    \label{fig:cht_setup}
    \vspace{.5cm}
    \end{subfigure}
    \begin{subfigure}[t]{.48\textwidth}
	\includegraphics[trim={0 -2.5cm 0 0},clip,width=\textwidth]{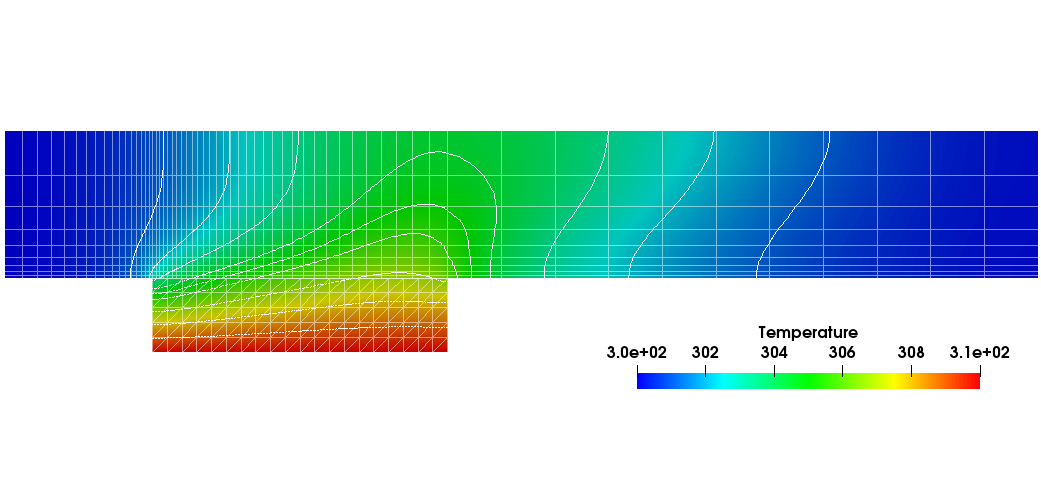}
	\caption{Steady-state temperature distribution. For better visualization, the depicted meshes are 5 times coarser than in reality.}
	\label{fig:cht_result}
    \end{subfigure}
    \hfill
    \begin{subfigure}[t]{.48\textwidth}
	\begin{tikzpicture}
\begin{axis}[
font=\footnotesize,
tuftelike,
x label style={at={(axis description cs:0.5,-0.3)}},
y label style={at={(axis description cs:-0.23,.5)}},
xlabel={x coordinate (m)}, 
ylabel={Temperature (K)}, 
align=center, 
xtick = {-0.5, 0, 0.5, 1, 1.5, 2, 2.5, 3},
legend style={at={(1,1.4)},anchor=north east, draw=none, align=left},
width=.9\textwidth
]

\addplot[color=blue] table[x=Points0 , y=T, col sep=comma]{images/plate/of-fe_linedata.csv};
\addlegendentry{OpenFOAM-FEniCS};

\addplot[color=red, dotted, dash pattern=on .001pt off 8pt, line cap=round, line width = 2pt] table[x=Points0 , y=T, col sep=comma]{images/plate/of-of_linedata.csv};
\addlegendentry{OpenFOAM-OpenFOAM};

\end{axis}
\end{tikzpicture}
	\caption{Steady-state temperature profile along a line $0.01 \text{m}$ above and parallel to the coupling interface $\Gamma_{C\text{,no-slip}}$. Both profiles differ by a relative l2 error of $7.61 \cdot 10^{-5}$.}
	\label{fig:cht_lineplot_compare}
	\end{subfigure}
\caption{Conjugate heat transfer testcase}
\end{figure}

As a simple CHT test case, we consider a flow over a heated plate\footnote{\url{https://github.com/precice/tutorials/tree/a166efa/flow-over-heated-plate}} inspired by \cite{VYNNYCKY199845}. Figure \ref{fig:cht_setup} sketches the domain and lists all physical parameters.
Heat conduction in the plate in the lower part of the domain is simulated with FEniCS, using a very similar application code as already explained in Section \ref{ssec:snippets}.
On top of the plate, we simulate an fluid flow from left to right with the OpenFOAM \cite{OpenFOAMJASAK200989} solver \texttt{buoyantPimpleFoam}. To couple OpenFOAM, we make use of the OpenFOAM-preCICE adapter\footnote{\url{https://github.com/precice/openfoam-adapter/tree/0898346}} \cite{Chourdakis2017}.
We use a Dirichlet-Neumann coupling: The solid participant (FEniCS) receives temperature values from the fluid participant and uses them as Dirichlet boundary condition at the coupling interface. The fluid participant (OpenFOAM), on the other hand, receives heat flux values from the solid participant and uses them as Neumann boundary condition. For the Dirichlet boundary condition in FEniCS, we use a FEniCS expression as described in Section \ref{ssec:SoftwareFunctionalities}.  
To map coupling data between non-matching meshes at the coupling interface, we use a nearest-neighbor mapping (in preCICE). 
Figure \ref{fig:cht_result} shows the temperature distribution of the coupled simulation after approaching steady-state. To verify our results, we compare the OpenFOAM-FEniCS coupling to an already existing OpenFOAM-OpenFOAM coupling. For comparable mesh resolutions, the results match very well, see Figure \ref{fig:cht_lineplot_compare}.

\subsection{Fluid-structure interaction with FEniCS and SU2}\label{fsi}

\begin{figure}
\begin{subfigure}{\textwidth}
\begin{minipage}[t]{.6\textwidth}
\begin{center}
\begin{tikzpicture}[scale=1, every node/.style={font=\small}]

\coordinate(origin) at (0,0);

\coordinate(lengthFlap) at (0,1);
\coordinate(thicknessFlap) at (0.1,0);

\coordinate(leftTopFlap) at ($(origin) + (0,-2em)$);
\coordinate(rightTopFlap) at ($(leftTopFlap)+(thicknessFlap)$);
\coordinate(leftBottomFlap) at ($(leftTopFlap)-(lengthFlap)$);
\coordinate(rightBottomFlap) at ($(rightTopFlap)-(lengthFlap)$);

\coordinate(channelWidth) at (0,4);
\coordinate(channelPreFlap) at (-2.95,0);
\coordinate(channelPostFlap) at (2.95,0);

\coordinate(leftBottomInflow) at ($(origin)+(channelPreFlap)$);
\coordinate(leftTopInflow) at ($(leftBottomInflow)+(channelWidth)$);

\coordinate(rightBottomOutflow) at ($(origin)+(thicknessFlap)+(channelPostFlap)$);
\coordinate(rightTopOutflow) at ($(rightBottomOutflow)+(channelWidth)$);

\draw[fill=participantBcolor!50](leftBottomInflow) -- (origin) -- ($(origin)+(lengthFlap)$) -- ($(origin)+(thicknessFlap)+(lengthFlap)$) -- ($(origin)+(thicknessFlap)$) -- (rightBottomOutflow) -- (rightTopOutflow) -- (leftTopInflow) -- cycle;
\node[] at ($.25*(leftBottomInflow)+.25*(leftTopInflow)+.25*(rightBottomOutflow)+.25*(rightTopOutflow)$){SU2 solves fluid problem};
\draw[fill=participantAcolor!50](leftTopFlap) -- (rightTopFlap) -- (rightBottomFlap) -- (leftBottomFlap) -- cycle;
\node[anchor=west, align=left](FEniCSLabel) at ($.25*(rightTopFlap)+.75*(rightBottomFlap)+(.5,-.2em)$){FEniCS solves\\ structure problem};

\draw[participantBcolor,very thick](leftBottomFlap) -- (leftTopFlap) -- (rightTopFlap) -- node[right]{$\Gamma_\text{C}$} (rightBottomFlap);  
\coordinate(midStructure) at ($.25*(leftBottomFlap)+.25*(leftTopFlap)+.25*(rightTopFlap)+.25*(rightBottomFlap)$);


\path [name path=A--B](FEniCSLabel.west) to[in=-60, out=180] (midStructure);
\path [name path=C--D](rightTopFlap) -- (rightBottomFlap);
\path [name intersections={of=A--B and C--D,by=E}];
\draw [->](FEniCSLabel.west) to[in=-50, out=180] (E);

\draw[participantAcolor,very thick](origin) -- ($(origin)+(lengthFlap)$) -- ($(origin)+(lengthFlap)+(thicknessFlap)$) -- node[right]{$\Gamma_\text{C}$} ($(origin)+(thicknessFlap)$);  
\draw(leftBottomFlap) -- node[below, black]{$\Gamma_\text{fixed}$} (rightBottomFlap);  

\draw(leftBottomInflow) -- node[above]{$\Gamma_\text{no-slip}$} (origin);
\draw(rightBottomOutflow) -- node[above]{$\Gamma_\text{no-slip}$} ($(origin)+(thicknessFlap)$); 
\draw(leftTopInflow) -- node[below]{$\Gamma_\text{no-slip}$} (rightTopOutflow);  
\draw(leftTopInflow) -- node[right]{$\Gamma_\text{inflow}$}(leftBottomInflow);  
\draw(rightBottomOutflow) -- node[left]{$\Gamma_\text{outflow}$}(rightTopOutflow);  


\dimline[extension start length=0, extension end length=0] {($(leftBottomInflow)+(-0.75em,0)$)}{($(leftTopInflow)+(-0.75em,0)$)}{$H$};
\dimline[extension start length=0, extension end length=0] {($(leftTopInflow)+(0,0.75em)$)}{($(rightTopOutflow)+(0,0.75em)$)}{$L$};
\dimline[extension start length=0, extension end length=0] {($(leftBottomInflow)-(0,0.75em)$)}{($(origin)-(0,0.75em)$)}{$l$};
\dimline[extension start length=0, extension end length=0, label style={fill=participantBcolor!50}] {($(origin)-(0.75em,0)$)}{($(origin)+(lengthFlap)-(0.75em,0)$)}{$h$};
\dimline[extension start length=0, extension end length=0, label style={fill=participantBcolor!50, above=0.6ex}, line style={arrows=dimline reverse-dimline reverse}] {($(origin)+(lengthFlap)+(0,0.75em)$)}{($(origin)+(lengthFlap)+(thicknessFlap)+(0,0.75em)$)}{$w$};
     
\foreach \i in {0,...,5}{%
      \draw[->,participantBcolor] ([yshift=\i * 0.2cm, xshift = -.5cm]leftBottomFlap) -- ([yshift=\i * .2cm]leftBottomFlap);}
      \path([xshift = -.25cm]leftTopFlap) -- node[above, participantBcolor]{$F$} (leftTopFlap);
      
\draw[<->] ($(origin)+(thicknessFlap)+.25*(lengthFlap)+(.1em,0)$) to[in=30, out=-30] node[right]{coupling via preCICE}($.25*(rightBottomFlap)+.75*(rightTopFlap)+(.1em,0)$);   

\end{tikzpicture}
\end{center}
\end{minipage}
\begin{scriptsize}
\begin{minipage}[t]{.4\textwidth}
\vspace{-6cm}
\begin{tabular}{cc}
\multicolumn{2}{c}{Dimensions}\\
\midrule
$L$ & $6.00\ \text{m}$\\
$H$ & $4.00\ \text{m}$\\
$l$ & $2.95\ \text{m}$\\
$h$ & $1.00\ \text{m}$\\
$w$ & $0.10\ \text{m}$\\[.5em]

\multicolumn{2}{c}{Fluid}\\
\midrule
$\text{Ma}_\infty$ & $0.01$\\  
$p_\infty$ & $101325\ \text{Pa}$\\ 
$T_\infty$ & $288.15\ K$ \\  
$\alpha$ (AOA) & $0\degree$ \\[.5em] 

\multicolumn{2}{c}{Solid}\\
\midrule
$E$ & $4\cdot10^6\ \text{N}/\text{m}^2$ \\  
$\nu_s$ & $3\cdot 10^{-1}$ \\  
$\rho_s$ & $3 \cdot 10^3\ \text{kg}/\text{m}^3$\\  
\end{tabular}
\vspace{-.2cm}
\end{minipage}
\end{scriptsize}
\vspace{-.3cm}
\caption{Geometric setup and physical parameters.}
\label{fig:fsi_setup}
\end{subfigure}

\begin{subfigure}[t]{\textwidth}
    \begin{center}
    \vspace{.1cm}
	\includegraphics[width=.5\linewidth]{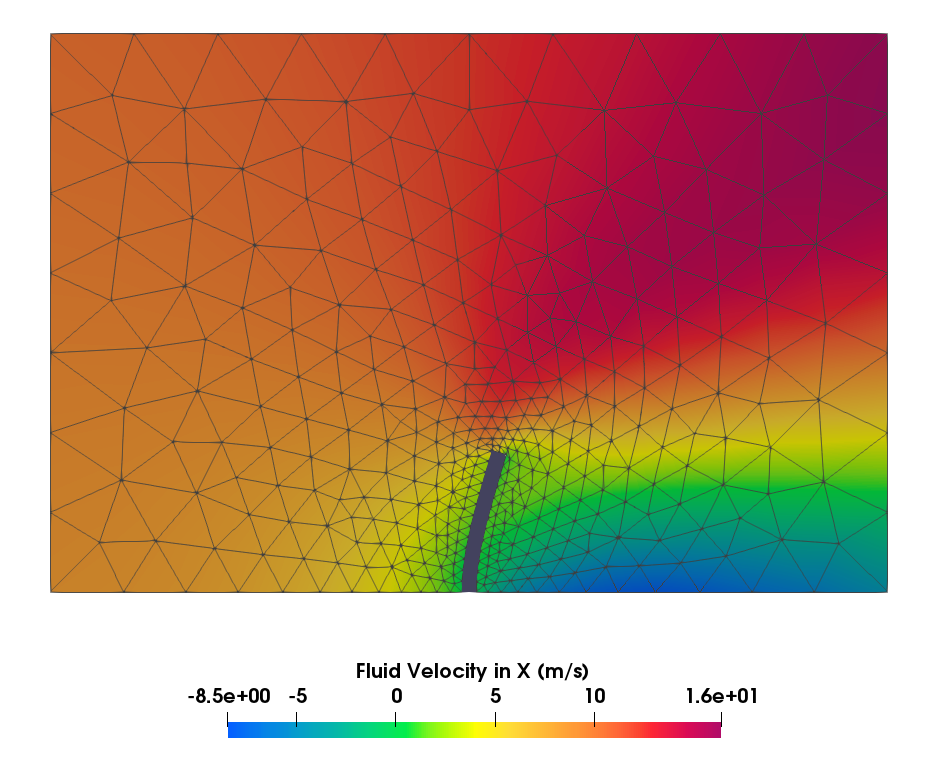}
    \end{center}
	\caption{Velocity field of fluid domain}
	\label{fig:fsi_fluid_result}
\end{subfigure}
\begin{center}
\begin{subfigure}{\textwidth}
    \begin{center}
  	\begin{tikzpicture}
\begin{axis}[
font=\footnotesize,
tuftelike,
x label style={at={(axis description cs:0.5,-0.3)}},
y label style={at={(axis description cs:-0.11,.5)}},
xlabel={Time (sec)}, 
ylabel={Tip Displacement (m)}, 
align=center, 
xtick = {0,1,2,3,4,5},
xtick = {0,1,2,3,4,5},
legend style={at={(1.17,1)},anchor=north east, draw=none, align=left}
]

\addplot[color=blue] table[x=time , y=tip_displ_fenics, col sep=comma]{images/fsi/compare_data_dealii_fenics_fsi.csv};
\addlegendentry{SU2-FEniCS};

\addplot[color=red, dotted, dash pattern=on .001pt off 8pt, line cap=round, line width = 2pt] table[x=time , y=tip_displ_dealii, col sep=comma]{images/fsi/compare_data_dealii_fenics_fsi.csv};
\addlegendentry{SU2-deal.II};

\end{axis}
\end{tikzpicture}
    \end{center}
    \vspace{-.5cm}
	\caption{Tip displacement in x direction of the elastic flap over time. Both curves differ by a relative l2 error of $8.17 \cdot 10^{-3}$.}
	\vspace{-.1cm}
	\label{fig:fsi_lineplot_compare}
\end{subfigure}
\end{center}
	\caption{Fluid-structure interaction test case}
\end{figure}
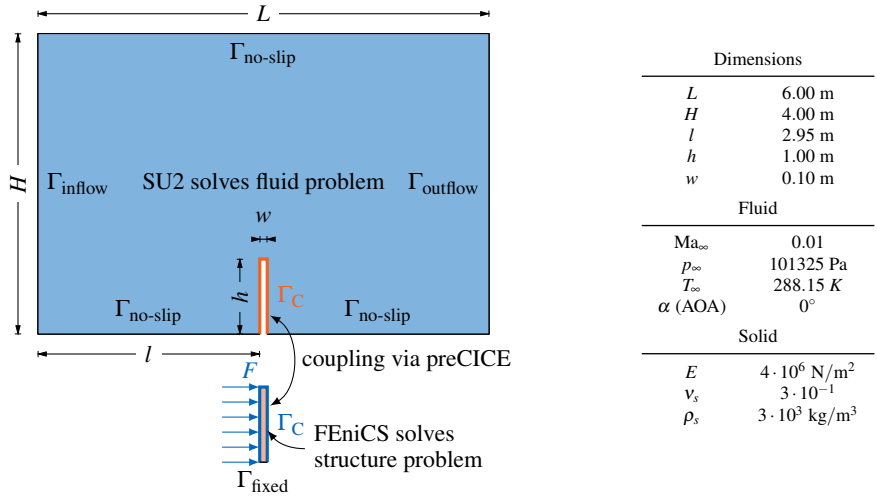
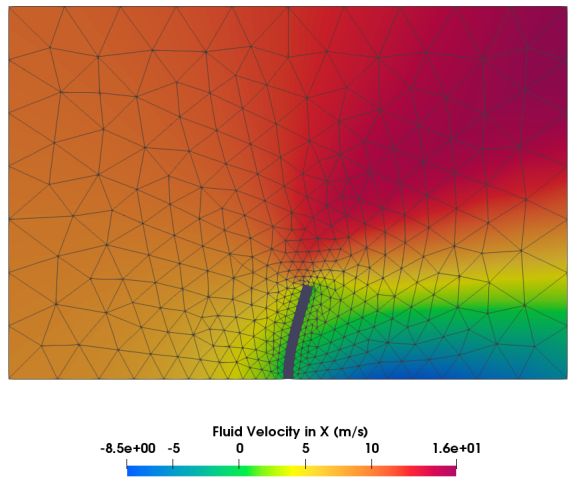
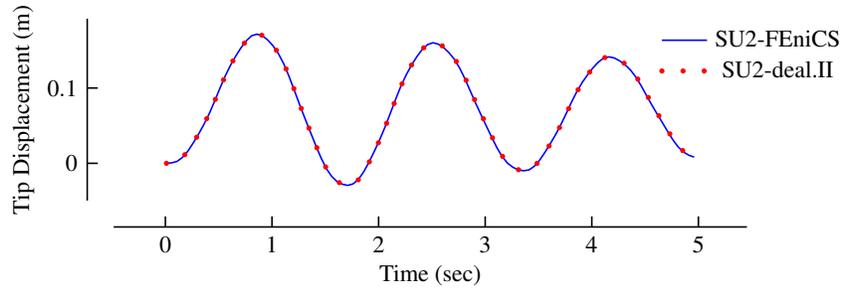

As a simple FSI test case, we consider a wall-mounted elastic flap in a channel flow\footnote{\url{https://github.com/precice/tutorials/tree/a166efa/perpendicular-flap}}. \autoref{fig:fsi_setup} sketches the domain and lists all physical parameters.
To simulate the elastic flap, we use a linear elasticity code in FEniCS, which was developed in the Bachelor thesis of Richard Hertrich \cite{Hertrich2019} following an example from \cite{jeremy_bleyer_2018_1287832}.
In the fluid domain, we use the compressible Euler solver of SU2 \cite{economon2016}. To couple SU2, we make use of the SU2-preCICE adapter\footnote{\url{https://github.com/precice/su2-adapter/tree/951159f}} \cite{Rusch2016}. 
We again use a Dirichlet-Neumann coupling: The fluid participant (SU2) receives displacement values from the fluid participant, computes velocity values from the displacement values, and uses the velocity values as Dirichlet boundary condition at the coupling interface. The solid participant (FEniCS), on the other hand, receives force values from the fluid participant and uses them as Neumann boundary condition. For the Neumann boundary condition in FEniCS, we now use point sources as described in Section \ref{ssec:SoftwareFunctionalities}.  
To map coupling data between non-matching meshes at the coupling interface, we again use a nearest-neighbor mapping (in preCICE). 
\autoref{fig:fsi_fluid_result} shows the fluid velocity at maximum deformation of the beam. To verify our results, we compare the SU2-FEniCS coupling to an already existing SU2-deal.II coupling. We use deal.II v9.2 \cite{dealII92} and the preCICE-deal.II adapter\footnote{\url{https://github.com/precice/dealii-adapter/tree/1d14846}}. For comparable mesh resolutions, the results match once again very well, see Figure \autoref{fig:fsi_lineplot_compare}.

\section{Impact}
\label{sec:impact}

The FEniCS-preCICE adapter enables the coupling of existing FEniCS application codes to other simulation software in only a few lines of code.
In particular, this holds for simulation software for which preCICE adapters already exist such as OpenFOAM, SU2, or deal.II.
Application scientists can now focus on coupled problems from the physical perspective and let the FEniCS-preCICE adapter handle the technical aspects: converting mesh and data structures, handling coupling conditions, or checkpointing for implicit coupling.   

The increase in opportunities works in both directions: Not only existing FEniCS users can now easily connect to the preCICE community and other simulation software, but also other communities (e.g.\ the large OpenFOAM community) can now directly benefit from FEniCS.

We want to illustrate the range of opportunities with three examples:
\begin{itemize}
  \item Within the collaborative research center 1313 \cite{SFB1313} several porous media applications are studied, such as the hydromechanical coupling of fractures and porous media. The current implementation is based on FEniCS and preCICE \cite{Schmidt2021} as it lowers the implementation hurdles compared to previous implementations based on DUNE \cite{Schmidt2019}. Furthermore, basing the implementation on FEniCS and preCICE immediately enables parallel computing capabilities. The current realization, however, uses the preCICE API directly from the application code. Using the new FEniCS-preCICE adapter instead will lead to a more idiomatic incorporation of preCICE into the coupled FEniCS codes allowing the researchers to rather focus on the implementation and evaluation of their models than the coupling itself. Some extensions of the adapter will be necessary, however, to tackle mixed-dimensional problems as they appear in hydromechanical coupling simulations.
  \item Together with researchers from Helmholtz-Zentrum Geesthacht and HSU Hamburg, we are currently investigating the coupling of electro-chemistry models to fracture mechanics to simulate corrosion\footnote{A prototype implementation is available at \url{https://github.com/uekerman/Coupled-Brittle-Fracture}}. For both fields, promising models \cite{Hoche2014_Corrosion} or codes in FEniCS \cite{Natarajan2019_Fenics_Fracture} exist. In contrast to both examples in Section \ref{sec:examples}, a volume coupling will be necessary. Since the adapter treats coupling conditions as general FEniCS \mintinline{python}{Expressions}, however, an extension to volume coupling will be straight forward.
  \item The layered design created by the FEniCS-preCICE adapter makes it also possible to easily prototype and test new numerical coupling algorithms, such as quasi-Newton waveform iteration in \cite{QNWI}. Here, an additional layer between the adapter and preCICE was used to implement and test a waveform data structure separately.   
\end{itemize}

In view of the rapid growth of the preCICE community and the popularity of FEniCS, we expect significant interest in the adapter in the next years. At this point in time (March 2021), preCICE is used by more than 80 research groups\footnote{Some preCICE users wrote testimonials on \url{https://www.precice.org}.}, spread over academia, non-academic research centers, and industry.

\section{Conclusions}
\label{sec:conclusions}
The new software FEniCS-preCICE allows to couple FEniCS application codes to other simulation software via preCICE.
Our motivation was to make such external coupling as easy as possible for existing FEniCS users.
To this end, we hid technical coupling complexity, such as parallel data structures, data conversion, or checkpointing, within the new middle software layer.
We were able to implement FEniCS-preCICE as a library and to follow a FEniCS-native style. 
This allows users of FEniCS to couple existing FEniCS application codes to other simulation software by only adding a few easy-to-understand lines of code.
We illustrated the potential by coupling 2D FEniCS application codes to OpenFOAM and SU2 to simulate conjugate heat transfer and fluid-structure interaction problems.
The design of the new software already envisions future extensions to other coupled problems (e.g. volume coupling) and to 3D scenarios. 
The impact of the new software should be significant. We already mentioned two projects that plan to use the software in the future: coupling fracture mechanics to porous media flow and coupling fracture mechanics to electro-chemistry models.
FEniCS itself is currently undergoing a major redesign named DOLFIN-X. Initial investigation showed that it should be possible to easily port FEniCS-preCICE to DOLFIN-X, which we want to realize in future work.

\section{Conflict of Interest}

We wish to confirm that there are no known conflicts of interest associated with this publication and there has been no significant financial support for this work that could have influenced its outcome.

\section*{Acknowledgements}

We thank the group of Claus Führer and Philipp Birken at Lund University for hosting Benjamin Rodenberg in Lund, where the first prototype of the adapter was developed. Additionally we thank Peter Meisrimel from Lund University for technical support and helpful discussions regarding FEniCS.

Benjamin Rodenberg's research stay in Lund was funded by SPPEXA, the DFG (Deutsche Forschungsgemeinschaft, German Research Foundation) Priority Program 1648 -- \textit{Software for Exascale Computing}.
The work of Benjamin Uekermann has been funded by the European Union's Horizon 2020 research and innovation program under the Marie Sklodowska-Curie grant agreement No 754462 and by the DFG Cluster of Excellence EXC 2075 \textit{Data-Integrated Simulation Science (SimTech)}.
The work of Ishaan Desai has been funded by the DFG project \textit{preDOM}, project number 391150578, and by \textit{SimTech}.
The work of Alexander Jaust has been funded by the DFG SFB 1313, project number 327154368.

\printbibliography

@article{Keyes2013_MultiPhysics,
title = {{Multiphysics simulations: Challenges and opportunities}},
author = {Keyes, David E. and McInnes, Lois C. and Woodward, Carol and Gropp, William and Myra, Eric and Pernice, Michael and others},
doi = {10.1177/1094342012468181},
journal = {International Journal of High Performance Computing Applications},
number = {1},
pages = {4--83},
volume = {27},
year = {2013}
}

@article{preCICE,
title = "{{preCICE}} – A fully parallel library for multi-physics surface coupling",
journal = "Computers \& Fluids",
volume = "141",
pages = "250 - 258",
year = "2016",
note = "Advances in Fluid-Structure Interaction",
issn = "0045-7930",
doi = "https://doi.org/10.1016/j.compfluid.2016.04.003",
author = "Hans-Joachim Bungartz and Florian Lindner and Bernhard Gatzhammer and Miriam Mehl and Klaudius Scheufele and Alexander Shukaev and Benjamin Uekermann"
}

@article{AlnaesBlechta2015a,
  title = {The {{FEniCS}} project version 1.5},
  author = {Martin S. Aln{\ae}s and Jan Blechta and Johan Hake and August Johansson and Benjamin Kehlet and Anders Logg and others},
  year = {2015},
  journal = {Archive of Numerical Software},
  volume = {3},
  number = {100},
  doi = {10.11588/ans.2015.100.20553},
  page = {9-23},
}

@book{LoggMardalEtAl2012,
  title = {Automated solution of differential equations by the finite element method},
  author = {Anders Logg and Kent-Andre Mardal and Garth N. Wells},
  year = {2012},
  publisher = {Springer},
  doi = {10.1007/978-3-642-23099-8},
  isbn = {978-3-642-23098-1},
}

@article{weller1998tensorial,
  title={A tensorial approach to computational continuum mechanics using object-oriented techniques},
  author={Weller, Henry G and Tabor, G and Jasak, Hrvoje and Fureby, C},
  journal={Computers in Physics},
  volume={12},
  number={6},
  pages={620--631},
  year={1998},
  doi={10.1063/1.168744},
  publisher={AIP Publishing}
}

@article{economon2016,
    author = {Economon, T. D. and Palacios, F. and Copeland, S. R. and Lukaczyk, T. W. and Alonso, J. J.},
    title = {{SU2}: An open-source suite for multiphysics Simulation and design},
    journal = {AIAA Journal},
    number = {3},
    pages = {828--846},
    volume = {54},
    year = {2016},
    doi = {10.2514/1.J053813}
}

@article{dealII92,
  title     = {The \texttt{deal.II} library, version 9.2},
  author    = {Daniel Arndt and Wolfgang Bangerth and Bruno Blais and
               Thomas C. Clevenger and Marc Fehling and Alexander V. Grayver and others},
  journal   = {Journal of Numerical Mathematics},
  publisher = {De Gruyter},
  year      = {2020},
  volume    = {28},
  number    = {3},
  pages     = {131-146},
  DOI       = {10.1515/jnma-2020-0043}
}

@inproceedings{Uekermann2017,
	author = "Benjamin Uekermann and  Hans-Joachim Bungartz and  Lucia Cheung Yau and  Gerasimos Chourdakis and  Alexander Rusch", 
	title = "Official preCICE adapters for standard open-source solvers",
	booktitle = "Proceedings of the 7th GACM Colloquium on Computational Mechanics for Young Scientists from Academia",
	year = "2017",
	month = "Oct",
	language = "en",
	url = "https://www.gacm2017.uni-stuttgart.de/registration/Upload/ExtendedAbstracts/ExtendedAbstract_0138.pdf",
}

@misc{2019FEniCSMixedDim, 
  author       = {Cécile Daversin Catty and
                  Marie E. Rognes},
  title        = {Mixed-dimensional coupled finite elements in 
                   {{FEniCS}}},
  month        = oct,
  year         = 2019,
  publisher    = {Zenodo},
  doi          = {10.5281/zenodo.3557448}
}

@InProceedings{FEniCS-HPC,
author="Hoffman, Johan
and Jansson, Johan
and Degirmenci, Niyazi Cem
and Sp{\"u}hler, Jeannette Hiromi
and Vilela De Abreu, Rodrigo
and Jansson, Niclas
and Larcher, Aur{\'e}lien",
editor="Di Napoli, Edoardo
and Hermanns, Marc-Andr{\'e}
and Iliev, Hristo
and Lintermann, Andreas
and Peyser, Alexander",
title="{{FEniCS-HPC}}: Coupled multiphysics in computational fluid dynamics",
booktitle="High-Performance Scientific Computing",
year="2017",
publisher="Springer International Publishing",
address="Cham",
pages="58--69",
isbn="978-3-319-53862-4",
doi={10.1007/978-3-319-53862-4_6}
}

@article{Bergersen2020,
author = {Bergersen, Aslak W and Slyngstad, Andreas and Gjertsen, Sebastian and Valen-Sendstad, Kristian},
doi = {10.21105/joss.02089},
journal = {The Journal of Open Source Software},
number = {50},
pages = {2089},
title = {{turtleFSI : A robust and monolithic FEniCS-based fluid-structure interaction solver}},
volume = {5},
year = {2020}
}

@article{Damiani2020,
author = {Damiani, Leonardo Hax and Kosakowski, Georg and Glaus, Martin A. and Churakov, Sergey V.},
doi = {10.1007/s10596-019-09919-3},
issn = {15731499},
journal = {Computational Geosciences},
number = {3},
pages = {1071--1085},
publisher = {Computational Geosciences},
title = {{A framework for reactive transport modeling using FEniCS–Reaktoro: governing equations and benchmarking results}},
volume = {24},
year = {2020}
}

@article{Massing2015,
author = {Massing, Andr{\"{i}}¿½ and Larson, Mats G. and Logg, Anders and Rognes, Marie E.},
doi = {10.2140/camcos.2015.10.97},
journal = {Communications in Applied Mathematics and Computational Science},
number = {2},
pages = {97--120},
title = {{A Nitsche-based cut finite element method for a fluid-structure interaction problem}},
volume = {10},
year = {2015}
}

@article{behnel2011cython,
author = {Behnel, Stefan and Bradshaw, Robert and Citro, Craig and Dalcin, Lisandro and Seljebotn, Dag Sverre and Smith, Kurt},
title = {Cython: The Best of Both Worlds},
year = {2011},
issue_date = {March 2011},
publisher = {IEEE Educational Activities Department},
address = {USA},
volume = {13},
number = {2},
doi = {10.1109/MCSE.2010.118},
journal={Computing in Science Engineering}, 
month = mar,
pages = {31–39},
numpages = {9}
}

@ARTICLE{NumPy,
  author={S. {van der Walt} and S. C. {Colbert} and G. {Varoquaux}},

  journal={Computing in Science Engineering}, 

  title={The {{NumPy}} array: A structure for efficient numerical computation}, 

  year={2011},

  volume={13},

  number={2},

  pages={22-30},

  doi={10.1109/MCSE.2011.37}}

@inproceedings{Lindner2017RBF,
title = {{Radial basis function interpolation for black-box multi-physics simulations}},
author = {Lindner, Florian and Mehl, Miriam and Uekermann, Benjamin},
booktitle = {Proceedings of the VII International Conference on Coupled Problems in Science and Engineering},
pages = {50--61},
publisher = {CIMNE},
url = {http://hdl.handle.net/2117/190255},
year = {2017}
}

@misc{SciPy,
author = {Jones, Eric and Oliphant, Travis and Peterson, Pearu and Others},
howpublished = {http://www.scipy.org/},
title = {SciPy: Open source scientific tools for Python},
year = {2001}
}

@phdthesis{Hertrich2019,
author = {Hertrich, Richard},
school = {Technical University of Munich},
title = {{Partitioned fluid structure interaction: Coupling FEniCS and OpenFOAM via preCICE}},
type = {Bachelor's Thesis},
year = {2019}
}

@article{VYNNYCKY199845,
title = {Forced convection heat transfer from a flat plate: the conjugate problem},
journal = {International Journal of Heat and Mass Transfer},
volume = {41},
number = {1},
pages = {45-59},
year = {1998},
issn = {0017-9310},
doi = {10.1016/S0017-9310(97)00113-0},
author = {M. Vynnycky and S. Kimura and K. Kanev and I. Pop}
}

@book{Langtangen2016,
author = {Langtangen, Hans Petter and Logg, Anders},
publisher = {Springer International Publishing},
address="Cham",
title = {Solving {{PDEs}} in {{Python}} - The {{FEniCS}} tutorial {{I}}},
doi = {10.1007/978-3-319-52462-7},
year = {2016}
}

@article{QNWI,
author = {Rüth, Benjamin and Uekermann, Benjamin and Mehl, Miriam and Birken, Philipp and Monge, Azahar and Bungartz, Hans-Joachim},
title = {Quasi-{N}ewton waveform iteration for partitioned surface-coupled multi-physics applications},
journal = {International Journal for Numerical Methods in Engineering},
volume = {n/a},
number = {n/a},
pages = {},
doi = {10.1002/nme.6443}
}

@article{fowler2007mocks,
  title={Mocks aren’t stubs},
  author={Fowler, Martin},
  url={http://martinfowler.com/articles/mocksArentStubs.html},
  year={2007}
}

@article{OpenFOAMJASAK200989,
	title = "{{OpenFOAM}}: Open source {{CFD}} in research and industry",
	journal = "International Journal of Naval Architecture and Ocean Engineering",
	volume = "1",
	number = "2",
	pages = "89 - 94",
	year = "2009",
	doi = "https://doi.org/10.2478/IJNAOE-2013-0011",
	author = "Hrvoje Jasak"
}

@phdthesis{Chourdakis2017,
author = {Chourdakis, Gerasimos},
school = {Technical University of Munich},
title = {A general {{OpenFOAM}} adapter for the coupling library {{preCICE}}},
type = {Master's Thesis},
year = {2017}
}

@article{Natarajan2019_Fenics_Fracture,
  title={A {{FEniCS}} implementation of the phase field method for quasi-static brittle fracture},
  author={Hirshikesh and Natarajan, Sundararajan and Annabattula, Ratna Kumar},
  journal={Frontiers of Structural and Civil Engineering},
  volume={13},
  pages={380--396},
  year={2019},
  doi={10.1007/s11709-018-0471-9}
}

@article{Hoche2014_Corrosion,
  title={Simulation of corrosion product deposit layer growth on bare magnesium galvanically coupled to aluminum},
  author={H{\"o}che, Daniel},
  journal = {Journal of The Electrochemical Society},
  pages = {C1--C11},
  volume={162},
  number={1},
  year={2014},
  publisher={IOP Publishing},
  doi={10.1149/2.0071501jes}
}

@phdthesis{Rusch2016 ,
	author = {Rusch, Alexander}, 
        school = {Technical University of Munich},	
	title = {Extending {{SU2}} to fluid-structure interaction via {{preCICE}}},
	type = {Bachelor's thesis},
	year = {2016}
}

@software{jeremy_bleyer_2018_1287832,
  author       = {Jeremy Bleyer},
  title        = {{Numerical tours of computational mechanics with FEniCS}},
  month        = jun,
  year         = 2018,
  publisher    = {Zenodo},
  doi          = {10.5281/zenodo.1287832},
}

@misc{Schmidt2021,
  author={Schmidt, Patrick and Jaust, Alexander and Steeb, Holger  and Mehl, Miriam},
  title={Simulation of flow in deformable fractures using a quasi-{N}ewton based partitioned coupling approach},
  howpublished = {In preparation},
}

@Article{Schmidt2019,
  author   = {Schmidt, Patrick and Steeb, Holger},
  title    = {Numerical aspects of hydro-mechanical coupling of fluid-filled fractures using hybrid-dimensional element formulations and non-conformal meshes},
  journal  = {GEM - International Journal on Geomathematics},
  year     = {2019},
  volume   = {10},
  number   = {1},
  pages    = {14},
  month    = {Feb},
  issn     = {1869-2680},
  day      = {10},
  doi      = {10.1007/s13137-019-0127-5},
}

@misc{SFB1313,
  title = {Sonderforschungsbereich 1313 homepage},
  howpublished = {\url{https://www.sfb1313.uni-stuttgart.de/}},
  note = {Accessed: 2020-12-23}
}

\end{document}